\documentclass[12pt]{iopart}

\usepackage{graphicx}
\begin{document}

\title[Correlated  fermions in lattices with spin-dependent disorder]{Thermodynamic properties of correlated  fermions in lattices with spin-dependent disorder}

\author{K. Makuch}
\address{Faculty of Physics, Institute of Theoretical Physics, University of Warsaw, ul.~Hoza~69,
00-681 Warszawa, Poland }
%\ead{Karol.Makuch@fuw.edu.pl}

\author{J. Skolimowski}
\address{Faculty of Physics, Institute of Theoretical Physics, University of Warsaw, ul.~Hoza~69,
00-681 Warszawa, Poland }
%\ead{Jan.Skolimowski@fuw.edu.pl}

\author{P. B. Chakraborty}
\address{Indian Statistical Institute, Chennai Centre, SETS Campus,
MGR Knowledge City, Taramani, Chennai 600113, India}
%\ead{prabuddha@isichennai.res.in}

\author{K. Byczuk}
\address{Faculty of Physics, Institute of Theoretical Physics, University of Warsaw, ul.~Hoza~69,
00-681 Warszawa, Poland}
%\ead{Krzysztof.Byczuk@fuw.edu.pl}

\author{D. Vollhardt}
\address{Theoretical Physics III, Center for Electronic Correlations and Magnetism,
Institute of Physics, University of Augsburg, D-86135, Augsburg, Germany
}
\ead{Dieter.Vollhardt@Physik.Uni-Augsburg.de}

\begin{abstract}
Motivated by the rapidly growing possibilities for experiments with ultracold atoms in optical lattices we investigate the thermodynamic properties of correlated lattice fermions in the presence of an external spin-dependent random potential.  The corresponding model, a Hubbard model with spin-dependent local random potentials, is solved within dynamical mean-field theory. This allows us to present a comprehensive picture of the thermodynamic properties of this system. In particular, we show that for a fixed total number of fermions spin-dependent disorder induces a magnetic polarization. The magnetic response of the polarized system differs from that of a system with conventional disorder.

\end{abstract}

\maketitle

\section{Introduction}

Interacting quantum many-particle systems in the presence of disorder are of great interest not only in condensed matter physics \cite{Mott90,Lee85,Altshuler85,Belitz94,Abrahams01,Abrahams10}
but recently also in the field of cold atoms in optical lattices
\cite{Lewenstein07,Fallani07,Billy08,Roati08,White09,Lewenstein2010,Kondov2011,Jendrzejewski12}.
In particular, ultracold gases have quickly developed into a fascinating new laboratory for the study of quantum many-body physics. A major advantage of cold atoms in optical lattices is the high degree of controlability of experimental parameters such as the interaction and the disorder strength, which allows for a detailed comparison between experiment and theory. Indeed, advanced techniques developed for experiments with ultracold gases in optical lattices facilitate the realization and explicit study of  model situations which are not accessible in solid state physics. For example, it was recently demonstrated that it is possible to prepare optical lattices with a \emph{spin-dependent} hopping of the quantum particles \cite{Mandel03, Mckay10,Soltan11}. Corresponding theoretical models had already been proposed earlier \cite{Liu04,Cazalilla05,Feiguin09,FK-model,FK,FK-RMP}.

In this paper we discuss the thermodynamic properties of interacting spin-1/2 lattice fermions with a repulsive, local interaction in the presence of a  spin-dependent random potential. In particular, we will compute the spin-dependent densities, the magnetization, the magnetic susceptibility, the compressibility, and  the spin-dependent densities of states.  When the interaction is taken to be attractive the model has a superfluid phase, which was recently investigated by Scalettar and collaborators \cite{Scalettar12}.  It may be realized experimentally in the following way: In an off-resonant electric field atoms experience
Stark shift of their ground state energies, which is proportional to the light intensity. The optical lattice is formed by superimposing laser waves propagating in different spatial directions. This effectively leads to a periodic  dependence of the electric field intensity and thereby to a periodic  potential.  Disorder can be introduced by focusing  a light beam on the optical lattice, which is scattered from a diffusive plate. Such an effective random potential (speckle-type disorder) is characterized by a very short correlation length and a pronounced statistical independence \cite{White09,Lewenstein2010,Pasienski2010}.  Using laser beams with different polarization
it is possible to generate a random potential which acts differently on  particles with different spin-orientation in the ground state \cite{Scalettar12}. The spin-dependent speckle disorder affects the  local one-particle energies, hopping amplitudes, and interparticle interaction potentials. Since ultracold atoms in optical lattices can be experimentally prepared in quite different ways, two model situations are of particular interest: (a) the total number of fermions is conserved, and (b) the number of fermions with different spin projection is individually conserved; the latter case is refered to as a ``spin-imbalanced'' system. Nanguneri \emph{et al.}~\cite{Scalettar12} analyzed case (b)  for an attractive interaction between the fermions, which  leads to a pairing instability. Employing Bogoliubov-de Gennes  mean-field theory they found  that spin-dependent disorder leads to a suppression of the pairing state.

In the first part of this paper we investigate the case when the total number of fermions is conserved (i.e., when the two spin populations have a common chemical potential) for a repulsive local interaction.  In particular, we will focus on fillings off half-filling, since in this case spin-dependent disorder induces a finite magnetization through the broadening of the spin sub-band on which the disorder acts. At the same time antiferromagnetism and characteristic correlation features which occur only at half-filling, such as the Mott metal-insulator transition \cite{Mott90,mott68,Gebhard,Georges,tokura,dmft_phys_today}, will be absent \cite{MIT-off-half-filling}. The spin-dependent randomness is introduced through local, random, and  spin-dependent potentials.
Such a model is closely related to electronic models investigated in solid state physics \cite{Gonis}, in particular  correlated lattice electrons in random potentials; see, for example, refs. \cite{Byczuk05,Byczuk09,Chakraborty11,Dobrosavljevic12,Miranda12}.
However, in solid state systems it is not possible (or, at least, not yet possible) to tune and manipulate spin-dependent potentials experimentally. By contrast, as discussed above, optical lattices with ultracold atoms do offer such possibilities. In particular, the model where the total number of fermions is conserved is applicable to systems with atoms characterized by a fast dipolar relaxation, such as Cr or rare elements with large magnetic dipole moments \cite{Fattori06}. Similarly, by employing double-shot phase contrast imaging for atoms in an external parabolic potential trap it is possible to determine spin-dependent densities, the spin susceptibility, and the compressibility of particles \cite{Ketterle13}.
Hence systematic studies of the disorder induced spin-asymmetric changes of the thermodynamic properties of such system are possible. In the second part of the paper we report results for spin-imbalanced systems, i.e., when each spin population has an individual chemical potential. Spin-imbalanced fermionic systems have been investigated experimentally \cite{Zwerlein06,Shin06,Navon10} and theoretically \cite{Parish07,Wunsch10,Snoek11,Gubbels12,Wolak12} either in optical lattices or in the case of a trap without disorder.
Spin-dependent disorder adds a new tool to the examination of correlation effects in spin-imbalanced fermionic systems.

\section{Hubbard model and spin-dependent disorder}

In the following we will study interacting fermions in a
random environment, modeled by the Anderson-Hubbard Hamiltonian with local,
spin-dependent disorder
\begin{equation}
\hat{H}=\sum_{ij,\sigma }t_{ij}\hat{c}_{i\sigma }^{\dagger }\hat{c}_{j\sigma} - \sum_{i\sigma} \mu_{\sigma} \hat{n}_{i\sigma} + \sum_{i\sigma}
\epsilon_{i\sigma} \hat{n}_{i\sigma} + U\sum_{i}\hat{n}_{i\uparrow }\hat{n}_{i\downarrow }.  \label{1}
\end{equation}
Here $\hat{c}_{\sigma}$ ($\hat{c}_{i\sigma }^{\dagger }$) is the anihilation (creation) fermionic operator, $\hat{n}_{i\sigma}=\hat{c}_{i\sigma }^{\dagger } \hat{c}_{i\sigma }$ is the particle number operator, $t_{ij}$ is the hopping matrix element, $U$ is the local interaction, and $\mu_{\sigma} $ is the chemical potential for particles with spin $\sigma$. In the model where the total number of fermions is conserved the chemical potentials of the two species obey  $\mu_{\sigma}= \mu_{-\sigma}= \mu$.
In the model where the numbers of fermions with different spin are individually conserved the chemical potentials $\mu_{\sigma}$ can differ.

The disorder is represented by random local potentials $\epsilon _{i\sigma}$ which are distributed according to a box probability density $P(\epsilon_{i\sigma})=1/\Delta_{\sigma}$ for $|\epsilon_{i\sigma}|\leq \Delta_{\sigma}/2$, and zero otherwise (``continuous disorder''). Here $\Delta_{\sigma} $ is the maximal energy difference between the local energies for a given spin direction and thus provides a measure of the disorder strength.
Averages over the disorder are
calculated by $\langle \cdots \rangle _{\mathrm{dis}}= \prod_{\sigma} \int d\epsilon_{\sigma}
P(\epsilon_{\sigma} )(\cdots )$, whereby the study of Anderson localization within the one-particle Green function formalism is excluded \cite{Lloyd,Thouless,Wegner}.

It is interesting to note \cite{Scalettar12,Scalettar_private} that by introducing the $z$-component of the magnetization operator
$\hat{m}_i=\hat{n}_{i\uparrow}-\hat{n}_{i\downarrow}$ and the  local particle number operator $\hat{n}_i=\hat{n}_{i\uparrow}+\hat{n}_{i\downarrow}$, the Anderson-Hubbard Hamiltonian for the model where the total number of fermions is conserved takes the form
\begin{equation}
\hat{H}=\sum_{ij,\sigma }t_{ij}\hat{c}_{i\sigma }^{\dagger }\hat{c}_{j\sigma} - \sum_{i}\mu_i^* \hat{n}_{i} + \sum_i h_i^* \hat{m}_i
+ U\sum_{i}\hat{n}_{i\uparrow }\hat{n}_{i\downarrow } ,  \label{2}
\end{equation}
where $\mu_i^*=\mu-(\epsilon_{i\uparrow} + \epsilon_{i\downarrow})/2$ and $h_i^*=(\epsilon_{i\uparrow} -
 \epsilon_{i\downarrow})/2$.
The Hamiltonian (\ref{2}) has a natural interpretation: it describes fermionic atoms moving in a
random chemical potential $\mu^*_i$ and a random Zeeman magnetic field $h_i^*$ which are correlated, since
\begin{equation}
\langle \mu_i^* h_j^* \rangle = \frac{1}{48}(\Delta_{\downarrow}^2-\Delta_{\uparrow}^2)\delta_{ij}.
\end{equation}
When $\Delta_{\downarrow}=\Delta_{\uparrow}$ we arrive at
the Anderson-Hubbard model with spin-independent disorder \cite{Byczuk-review}.

\section{Dynamical mean-field theory}
\label{DMFT}

We solve the Hamiltonian (\ref{1}) within dynamical mean-field theory  (DMFT) \cite{Metzner,Georges}.
Hence  all local dynamical correlations due to the local interaction are fully taken into account.
However, non-local correlations in position space are absent. Long-range order in space can be
included within the DMFT, but this will not be the case in the present investigation.

In the DMFT scheme the local Green function $G_{\sigma n}$ is determined by the
bare density of states (DOS) $N^{0}(\epsilon )$ and the local self-energy
$\Sigma _{\sigma n}$ as $G_{\sigma n}=\int d\epsilon N^{0}(\epsilon
)/(i\omega _{n}+\mu_{\sigma} -\Sigma _{\sigma n}-\epsilon )$. Here the
subscript $n$ refers to the Matsubara frequency $i\omega
_{n}=i(2n+1)\pi /\beta $ for the temperature $T$, with $\beta
=1/T$. Within DMFT the
local Green function $G_{\sigma n}$ is determined
self-consistently by
\begin{equation} G_{\sigma n}=- \langle\langle c_{\sigma n}c_{\sigma n}^{\star } \rangle \rangle_{\rm dis}
=-\Bigg\langle
\frac{\int D\left[ c_{\sigma },c_{\sigma }^{\star }\right]
c_{\sigma n}c_{\sigma n}^{\star }e^{{\cal A}_i\{c_{\sigma
},c_{\sigma }^{\star }, {\cal G}_{\sigma }^{-1}\}}}{\int D\left[
c_{\sigma },c_{\sigma }^{\star } \right] e^{{\cal A}_i\{c_{\sigma
},c_{\sigma }^{\star },{\cal G}_{\sigma }^{-1}\}}} \Bigg\rangle
_{\rm dis}, \label{3}
\end{equation}
together with the \textbf{k}-integrated Dyson equation
$\mathcal{G}_{\sigma n}^{-1}=G_{\sigma n}^{-1}+\Sigma _{\sigma
n}$ \cite{Georges,Ulmke}. The single-site action $\mathcal{A}_{i}$ for a site with the
ionic energy $\epsilon _{i} \in [- \frac{\Delta}{2}, \frac{\Delta}{2}] $ has the form
\begin{eqnarray}
\mathcal{A}_{i}\{c_{\sigma },c_{\sigma }^{\star },\mathcal{G}_{\sigma
}^{-1}\} =\sum_{n,\sigma }c_{\sigma n}^{\star }\mathcal{G}_{\sigma
n}^{-1}c_{\sigma n}-\sum_{\sigma }\int_{0}^{\beta }d\tau
\epsilon _{i\sigma} n_{\sigma }(\tau )  \nonumber \\
-\frac{U}{2}\sum_{\sigma }\int_{0}^{\beta }d\tau c_{\sigma }^{\ast }(\tau
)c_{\sigma }(\tau )c_{-\sigma }^{\ast }(\tau )c_{-\sigma }(\tau ),  \label{4}
\end{eqnarray}
where we used a mixed time/frequency representation for Grassmann variables $c_{\sigma }$,
$c_{\sigma }^{\star }$. In Eq.~\ref{3} the average over the quantum states with Boltzmann distribution (inner bracket) and over the disorder (outer bracket) is taken.  In the non-interacting case ($U=0$) the DMFT equations reduce to those of the coherent potential approximation \cite{Janis91,Vlaming,Vollhardt04}.

In the following we employ a semicircular model-DOS $N^{0}(\epsilon
)= (8/\pi W^2) \sqrt{W^2/4-\epsilon^2}$.
%, which corresponds to nearest-neighbor hopping on a Bethe lattice with infinite coordination number ($Z=\infty$). 
The half bandwidth $W=2$ (corresponding to a hopping amplitude  $t=1/2$) is taken as the unit of energy. 
%For a finite- dimensional Bravais lattice one has $W=Z t$.
%The DMFT with semicircular DOS is known to reproduce typical results for finite-dimensional lattices with the same band width very well
%\cite{Georges}.
The one-particle Green function
in Eq.~(\ref{3}) is calculated by solving the DMFT equations iteratively
\cite{ulmke98,wahle98} using Hirsch-Fye Quantum Monte-Carlo (QMC) simulations \cite{hirsh86}.

We now discuss and compare results for different types of disorder: i) \emph{spin-independent disorder}, where $\Delta_{\downarrow}=\Delta_{\uparrow}=\Delta$, and ii) \emph{spin-dependent disorder} with  $\Delta_{\downarrow}=\Delta$ and $\Delta_{\uparrow}=0$. In both cases $\Delta$ is a measure of the disorder strength. They represent extreme situations and thereby illustrate the most important differences between spin-independent and spin-dependent disorder. Intermediate cases can be studied in the same way.

In the model where the total number of fermions is conserved the chemical potential $\mu$ is fixed to keep the total number of fermions per site constant:
$n=n_{\uparrow}+n_{\downarrow}$, where $n_{\sigma} = \langle \langle \sum_i \hat{n}_{i\sigma}\rangle\rangle_{\rm dis} / N_L$ and $N_L$ is the number of lattice sites. The numerical investigations were performed for the densities $n=0.3$, $0.5$ and $0.7$ \cite{error-bars}. Qualitatively, all three cases lead to similar results. For the purpose of illustration we present  results only for $n=0.5$.

In the model where the numbers of particles with different spin $\sigma$ are separately conserved the chemical potentials $\mu_{\sigma}$ are fixed individually to keep the number of fermions with spin $\sigma$ per site constant. The numerical investigations were performed for $(n_{\uparrow},n_{\downarrow})=(0.2,0.3)$ , $(0.3,0.2)$, $(0.2,0.5)$, and $(0.5,0.2)$. For the purpose of illustration we here present results for $(0.5,0.2)$  since qualitatively, all cases lead to similar thermodynamic properties.

\section{Constant total number of fermions}

\subsection{Finite magnetization caused by spin-dependent disorder}

\begin{figure}[!]
\centerline{\includegraphics[width=12cm]{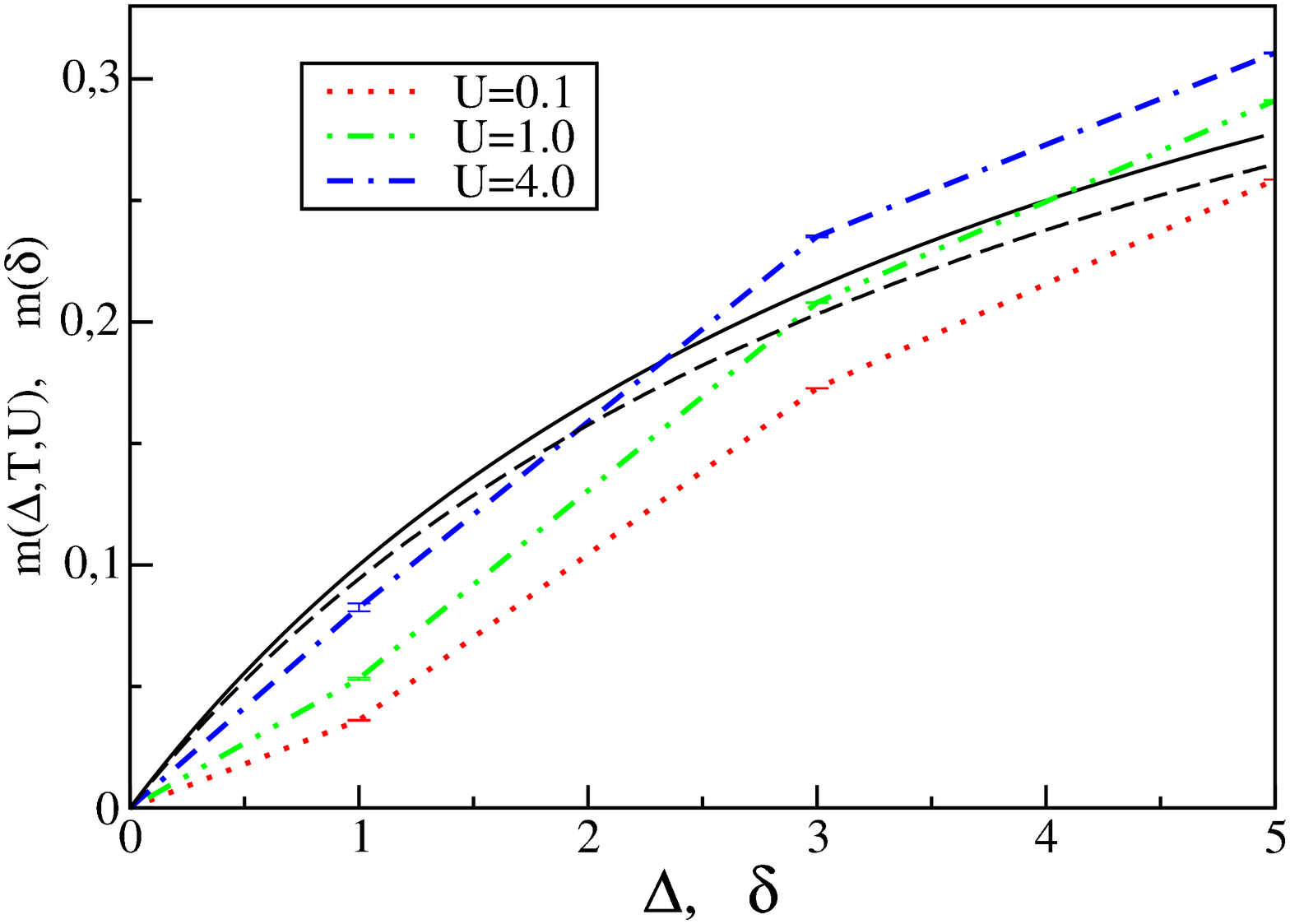}}
\centerline{\includegraphics[width=6cm]{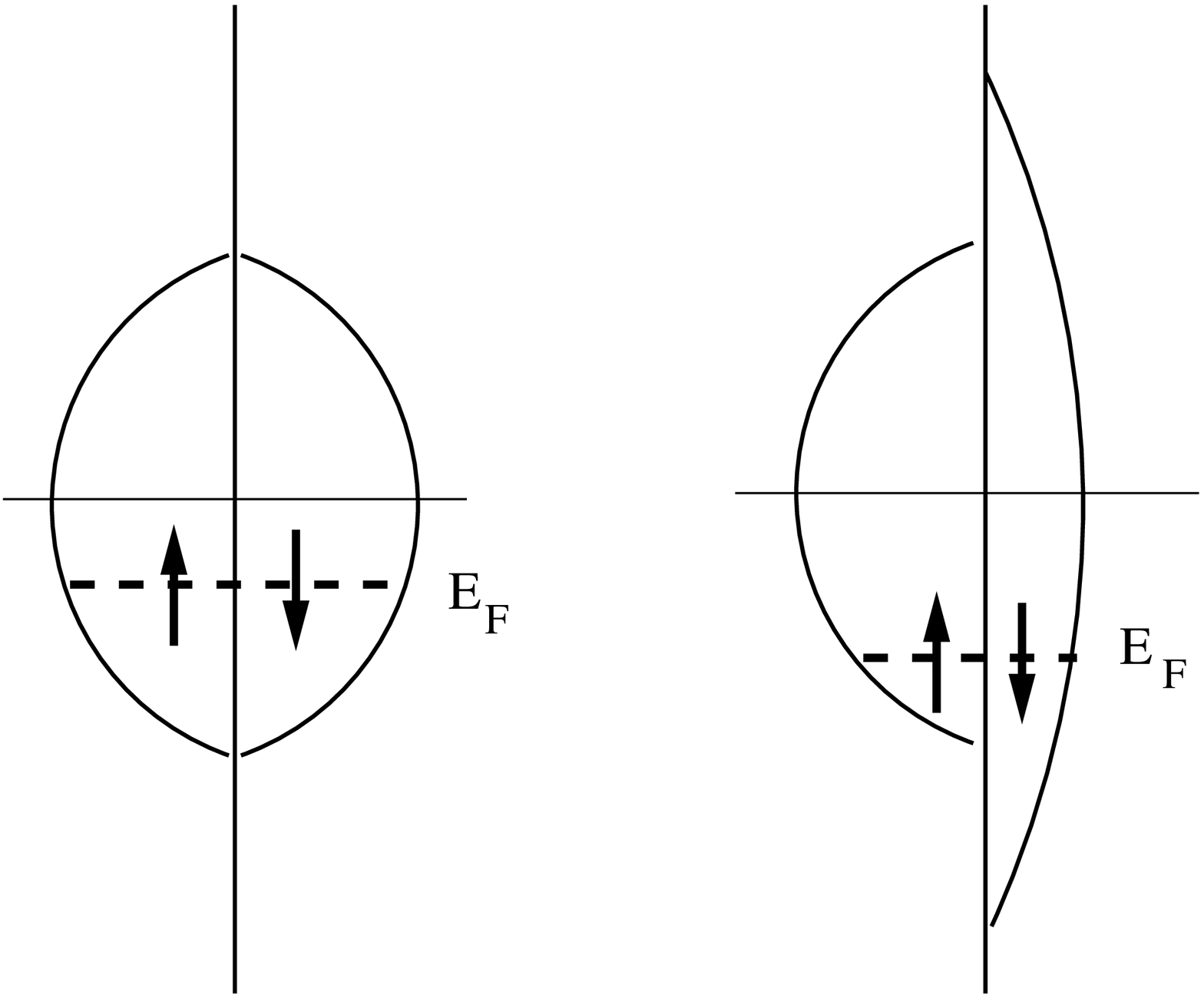}}
\caption{Constant total number of fermions:
Upper panel: Increase of the magnetization $m$ as a function of spin-dependent disorder $\Delta$ for interactions $U=0.1, 1.0, 4.0$ at $T=0.06$. Similar results are obtained from a non-interacting model where the band widths of particles with opposite spin differ by an energy $\delta$ (see text).  Lower panel: Schematic drawing of the influence of spin-dependent disorder. The disorder leads to a broadening of the respective spin subband. Due to the conservation of the total number of fermions a finite magnetization is built up. }
\label{fig2}
\end{figure}

The most striking effect caused by spin-dependent disorder is the appearance of a finite magnetization $m(\Delta,T,U)= \langle \langle \sum_i \hat{m}_{i}\rangle\rangle_{\rm dis} / N_L$ for $\Delta>0$ at arbitrary temperatures $T$ and interaction values $U$, as shown in the upper panel of Fig.~\ref{fig2} for the lowest temperature studied here, $T=0.06$. The magnetization $m$ is seen to grow monotonously as the spin-dependent disorder strength $\Delta$ is increased.
The increase of $m$ is slightly stronger for larger values of the interaction $U$.
It should be noted that the origin of the magnetization
 is not due to electronic correlations, but is a pure one-particle effect, which for a symmetric DOS, occurs only away from half filling ($n\neq 1$).
Namely, spin-dependent disorder leads to a symmetric broadening of the spin subband on which it acts (here the down-spins), whereby the number of spin-down particles \emph{increases} while the number of spin-up particles \emph{decreases} since the total number of particles remains constant (that is, the two spin populations have a
common chemical potential as shown  in the lower part of Fig.~\ref{fig2}). This then leads to a finite magnetization. Here we assumed that the system is able to relax to the new equilibrium state by means of spin-flip processes which are provided, for example, by the spin-orbit interaction or inelastic dipolar collisions \cite{Fattori06}, but which are not explicitly included in the model Hamiltonian (\ref{1}).  The weak influence of $U$ on the magnetization may be attributed to the flow of spectral weight induced by the correlations.

For a symmetric DOS and a half-filled band
the number of states below the Fermi energy $E_F=0$ remains constant when $\Delta$ is increased; this holds for any interaction $U$.
The magnetization is then zero.  This effect was clearly observed in our numerical simulations, but is not presented here.

The increase of the magnetization $m$ with increasing spin-dependent disorder $\Delta$  may be understood already within a simple model without interactions ($U=0$) where the particles with different spin orientation have different band widths \cite{Skolimowski,Hirsch,Kim}.
Namely, if we assume that the DOS of the system is  given by
\begin{equation}
N^0_{\sigma}(\epsilon) = \left\{
\begin{array}{ccc}
\frac{1}{W_{\sigma}} & {\rm for} & |\epsilon| \leq \frac{W_{\sigma}}{2},\\
0 & {\rm for} & |\epsilon| > \frac{W_{\sigma}}{2},
\end{array}
\right.
\end{equation}
where $W_{\uparrow}=W$ and $W_{\downarrow}=W+\delta$, the magnetization at $T=0$ is given by
\begin{equation}
m(n,\delta) = (1-n) \frac{\delta}{2W+\delta}.
\end{equation}
The increase of the magnetization as a function of the difference of the band widths,  $\delta$,  at $T=0$ is shown by the full black curve in the upper panel of Fig.~\ref{fig2}. A very similar result, indicated by a black dashed curve in the same figure, is obtained for a semi-elliptic DOS with spin-dependent band widths $N^{0}_{\sigma}(\epsilon
)= (8/\pi W_{\sigma}^2) \sqrt{W_{\sigma}^2/4-\epsilon^2}$.  Both curves closely resemble those calculated for the Hamiltonian (\ref{1}), except for the turning point observed in $m(\Delta,T,U)$ for $U=0.1, 1.0$. Apparently this feature is due to the scattering of the electrons by the random inhomogeneities, which is included in  the  frequency dependent self-energy of the local Green function.

\subsection{Thermodynamic properties}

\begin{figure}[!]
\centerline{\includegraphics[width=12cm]{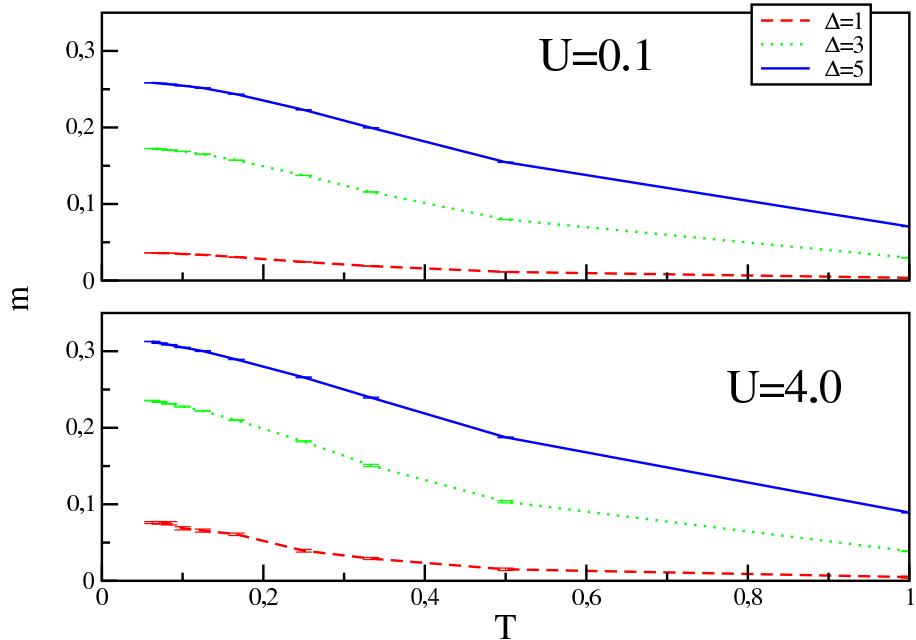}}
\caption{Constant total number of fermions: Magnetization induced by spin-dependent disorder as a function of temperature $T$  for different values of the interaction $U$ and disorder strength $\Delta$. }
\label{fig1}
\end{figure}

As the temperature increases the Fermi-Dirac distribution is smeared out and the arguments for the appearance of a finite magnetization due to a spin-dependent band broadening at a constant chemical potential, although still valid, become less stringent. Indeed, as shown in Fig.~\ref{fig1} the magnetization slowly decreases with increasing temperature {\bf as $1/T$} for all values of the interaction $U$ and disorder strengths $\Delta>0$.

\begin{figure}[h]
\centerline{\includegraphics[width=15cm]{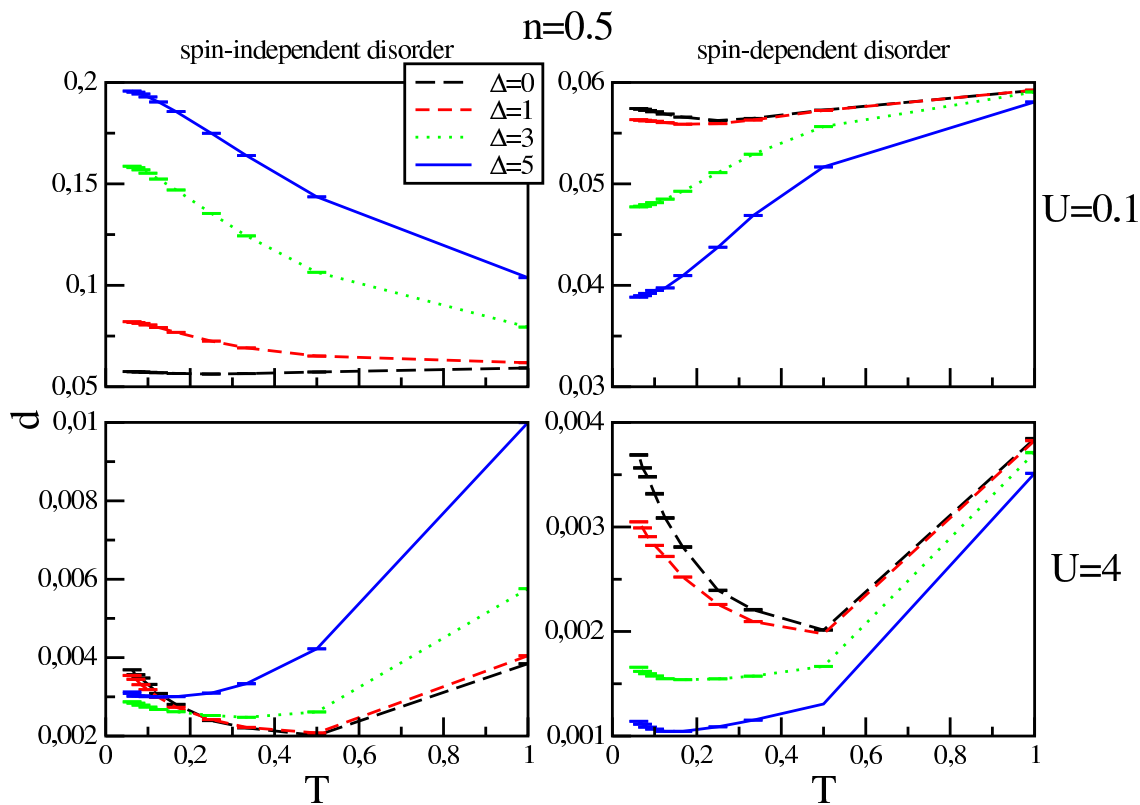}}
\caption{Constant total number of fermions: Double occupation $d$ as a function of the temperature $T$ for different values ($\Delta=0,1,3,5$) of spin-independent disorder (left) and spin-dependent disorder (right); upper panel: $U=0.1$, lower panel: $U=4$.}
\label{fig3}
\end{figure}

The magnetization induced by the disorder influences the thermodynamic properties of the system. In Fig.~\ref{fig3} we compare the temperature dependences of the average double occupation $d=\sum_i \langle\langle \hat{n}_{i\uparrow} \hat{n}_{i\downarrow} \rangle\rangle_{\rm dis}/N_L$ for spin-independent disorder (left panel) and spin-dependent disorder (right panel), respectively. For spin-independent disorder and weak interactions, e.g. $U=0.1$, an increase of the disorder $\Delta$  leads to a corresponding increase of $d$, i.e., on average more energy levels below the Fermi energy are occupied by two fermions with opposite spins.  For increasing $U$ a minimum appears in $d(T)$
 which corresponds to a maximum in the local moment per site $\mathcal{S}=\sqrt{\langle \sum_i \langle \hat{m}_i^2\rangle \rangle_{\rm dis}/N_L}=\sqrt{n-2d}$.
This is seen clearly for $U=4$ in Fig.~\ref{fig3}. Increasing the disorder $\Delta$ reduces the value of the local moment even in the presence of strong interactions ($U=4$), when $d$ is very small. Only at very low temperatures and $U=4$ we can see an opposite trend.

In the case of spin-dependent disorder (right panel of Fig.~\ref{fig3}) the double occupation is always found to be suppressed by the disorder $\Delta$.  This is an effect of the finite magnetization of the system. As a consequence the local moment increases. We also see that the local minimum of $d$ is shifted to lower temperatures when $\Delta$ is increased. Whether it disappears at $T=0$ cannot be decided here since we are not able to run QMC simulations at lower temperatures.

\begin{figure}[!]
\centerline{\includegraphics[width=15cm]{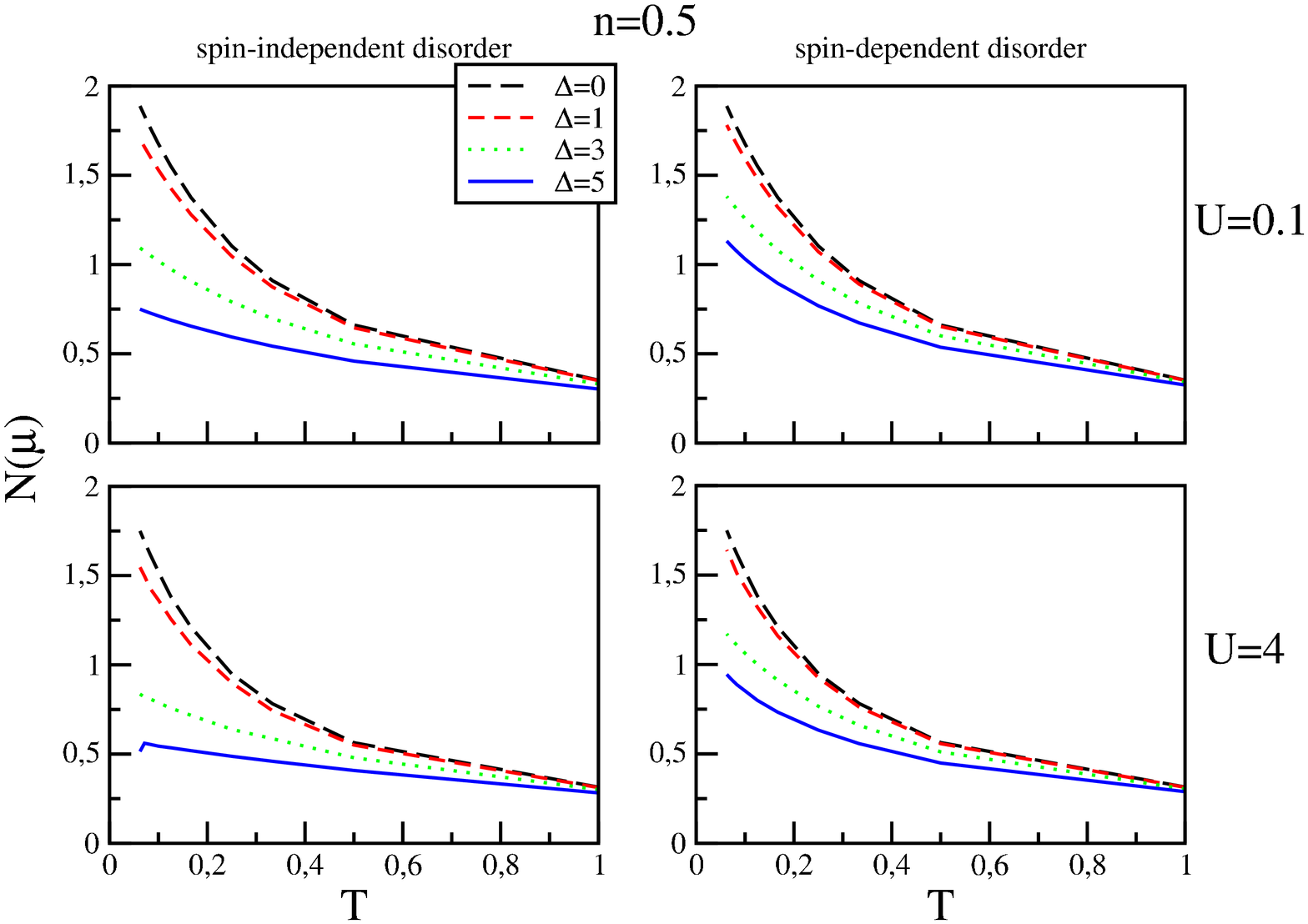}}
\caption{Constant total number of fermions: Density of states at the chemical potential as a function of the temperature $T$ for different values ($\Delta=0,1,3,5$) of spin-independent disorder (left) and spin-dependent disorder (right); upper panel: $U=0.1$,  lower panel: $U=4$.}
\label{fig4}
\end{figure}

Next  we compare the value of the DOS at the chemical potential $N(\mu)$ for different disorder strengths (Fig.~\ref{fig4}). As expected, an increase of the disorder reduces the DOS. We also see that this decrease is stronger for spin-independent disorder because the random potential acts equally on both spin subbands. When the temperature is lowered, the DOS increases similarly in both cases due to the enhanced quantum coherence of the system.

%\frac{m(T)}{dh}

Static susceptibilities provide useful information about possible phase instabilities of the system and their response to external fields. Therefore, we now compute
\begin{itemize}
\item the ferromagnetic (FM) susceptibility $\chi_{FM}= \frac{d m}{dh} \bigg|_{h\rightarrow 0}$, where $m$ is the magnetization density in the presence of an external magnetic field $h$,
\item the antiferromagnetic (AFM) susceptibility $\chi_{AFM}= \frac{d m_{st}}{dh_{st}} \bigg|_{h_{st}\rightarrow 0}$, where $m_{st}=n_{A\uparrow}-n_{B\uparrow}$ is the staggered magnetization density on a bipartite lattice with nonequivalent sites $A$ and $B$ in the presence of a staggered magnetic field $h_{st}$, where $n_{A(B)}=\langle \langle \sum_{i\in A(B)} \hat{n}_{i\sigma} \rangle \rangle _{\rm dis}$, and
\item the density susceptibility (compressibility) $\chi_{c}= \frac{d n}{d\mu }$, where $n$ is the particle number density at  the chemical potential $\mu$.
\end{itemize}

\begin{figure}[!]
\centerline{\includegraphics[width=15cm]{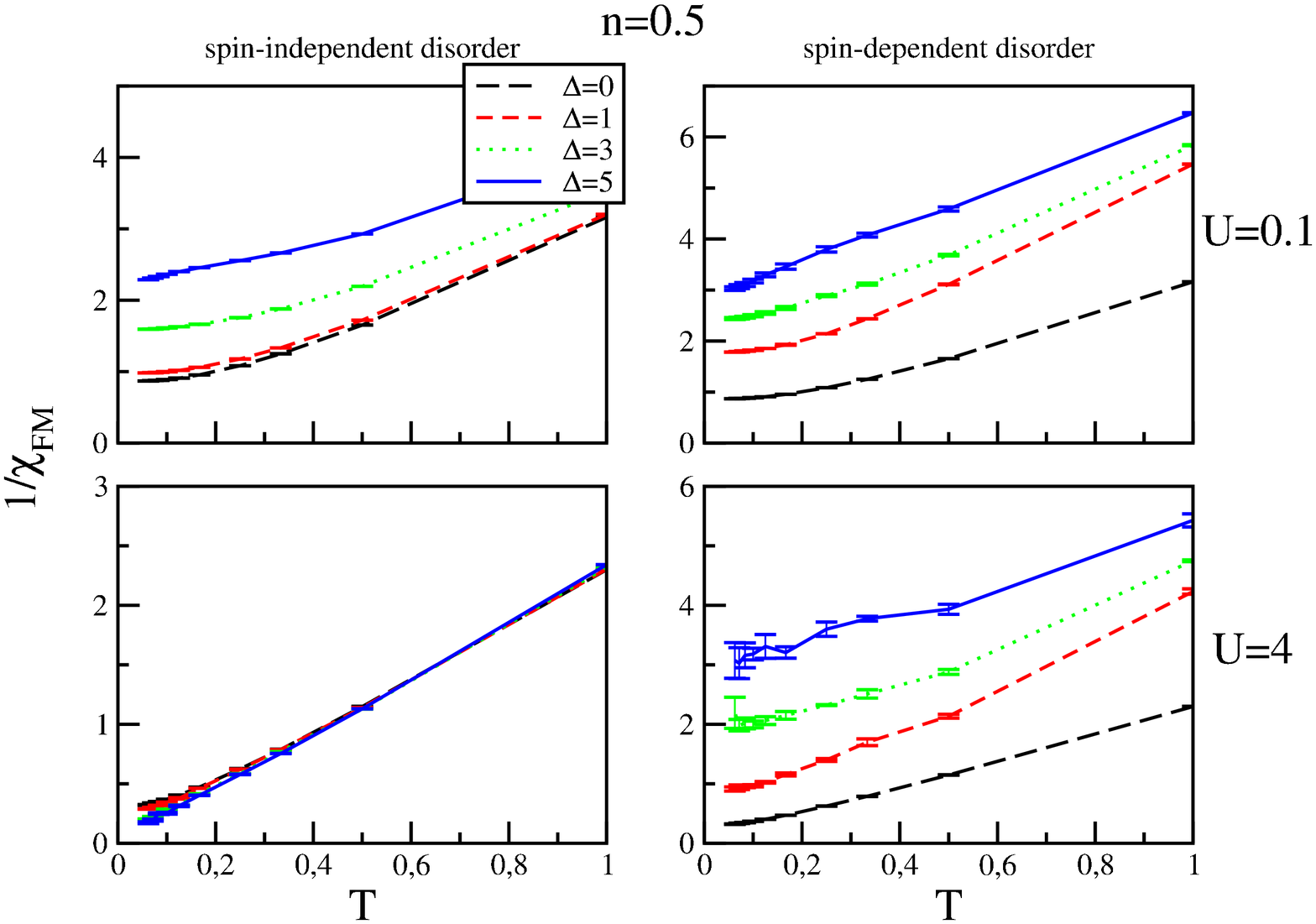}}
\caption{Constant total number of fermions: Inverse ferromagnetic susceptibility as a function of the temperature $T$ for different values ($\Delta=0,1,3,5$) of spin-independent disorder (left) and spin-dependent disorder (right); upper panel: $U=0.1$, lower panel: $U=4$.}
\label{fig5}
\end{figure}

\begin{figure}[!]
\centerline{\includegraphics[width=15cm]{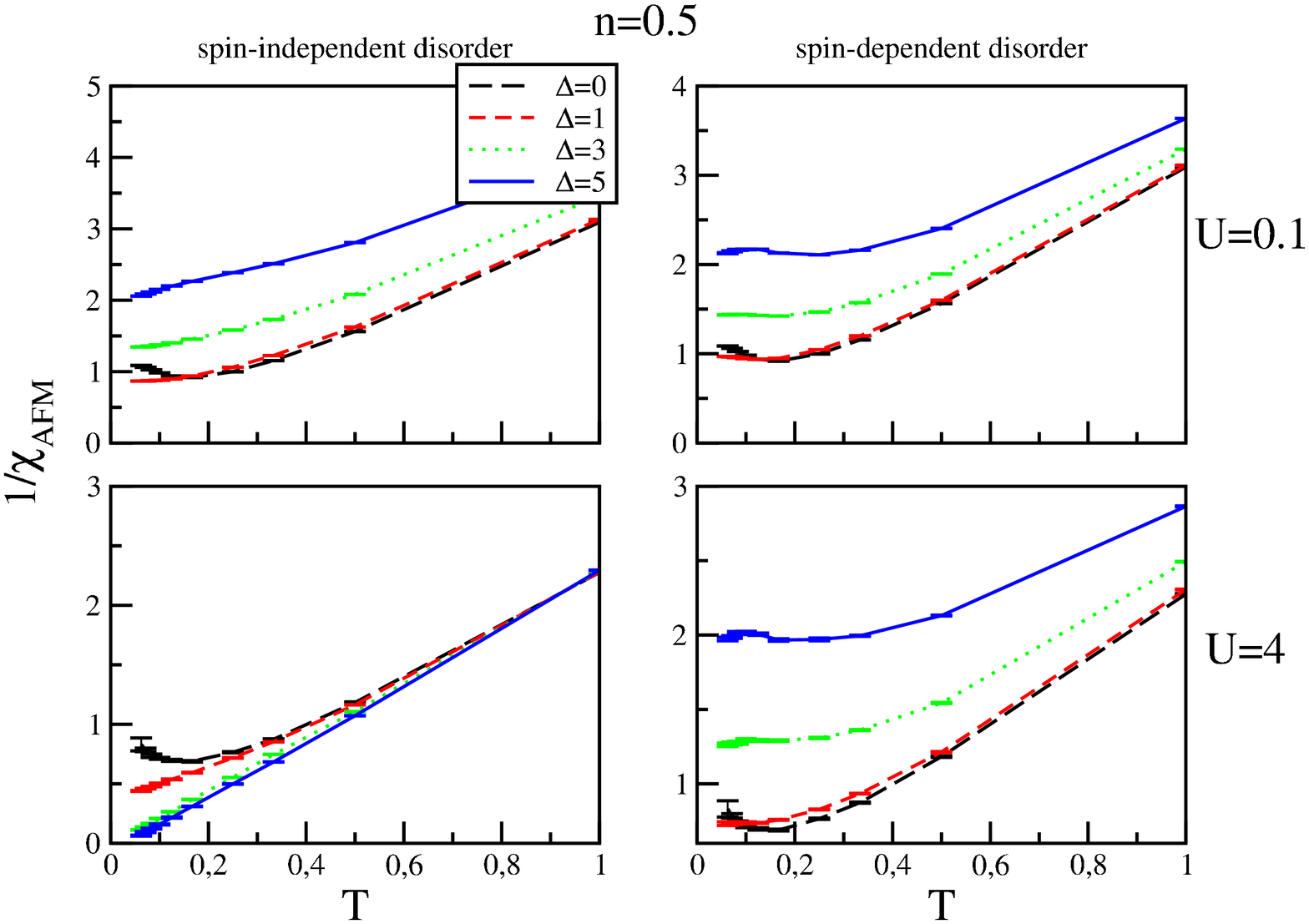}}
\caption{Constant total number of fermions: Inverse antiferromagnetic susceptibility as a function of the temperature $T$ for different values ($\Delta=0,1,3,5$) of spin-independent disorder (left) and spin-dependent disorder (right); upper panel: $U=0.1$, lower panel: $U=4$.}
\label{fig6}
\end{figure}

The magnetic response in the FM and AFM channels differ significantly for spin-independent and spin-dependent disorder, respectively: both susceptibilities are reduced by spin-independent disorder, at least for weak  interactions ($U=0.1$), as seen in the upper left panels of Figs.~\ref{fig5}~and~\ref{fig6}. For strong interactions in the presence of spin-independent disorder ($U=4$, lower left panels of Figs.~\ref{fig5}~and~\ref{fig6}) the FM and AFM susceptibilities behave  differently, in particular at  the lower temperatures. Namely, they exhibit Curie-like behavior, {\bf i.e. $\chi_{FM} \sim 1/T$,}  and increase with increasing disorder. This is due to the formation of local moments and corroborates the temperature behavior of the double occupation discussed earlier. Indeed, at low enough temperatures and for strong interactions the double occupation $d$ is reduced by spin-independent disorder.

 A similar Curie-like behavior of the FM susceptibility was previously found in the two-dimensional Hubbard model with random box potential \cite{Chakraborthy11_bis}. However, in this case neither the AFM susceptibility nor the averaged local moment show any unusual behavior. Therefore we conclude that in contrast to high dimensions, where correlated electrons in random potentials form local moments, the behavior in two dimensions is different and is yet unexplained.

In the case of spin-dependent disorder an  increase of the disorder strength $\Delta$ reduces both magnetic susceptibilities (see the right panels of Figs.~\ref{fig5}~and~\ref{fig6}). Namely, due to the magnetic polarization induced by the spin-dependent disorder the response of the system to a magnetic field is weaker.

\begin{figure}[!]
\centerline{\includegraphics[width=15cm]{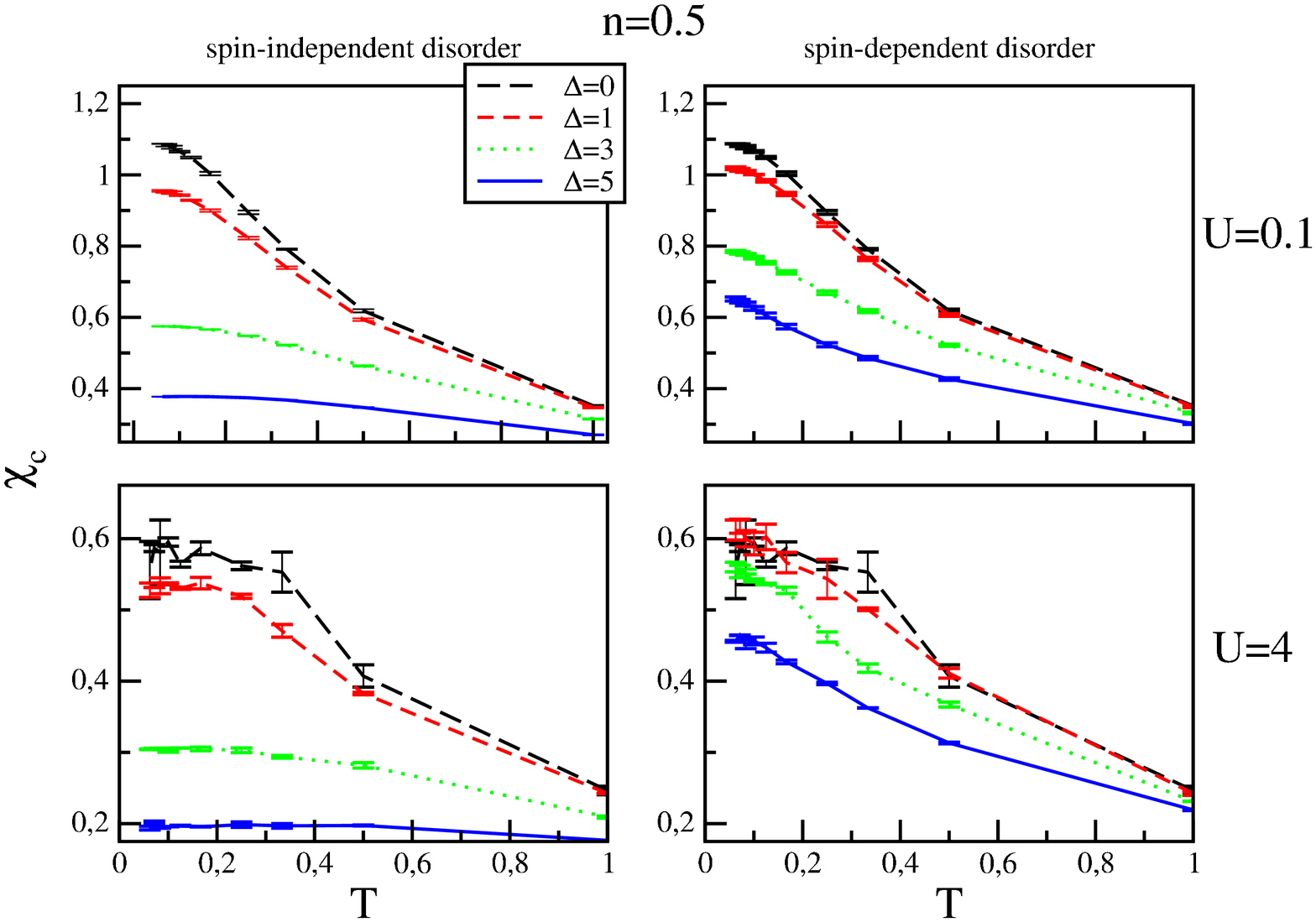}}
\caption{Constant total number of fermions: Compressibility $\chi_c$ as a function of the temperature $T$ for different values ($\Delta=0,1,3,5$) of spin-independent disorder (left) and spin-dependent disorder (right); upper panel: $U=0.1$, lower panel: $U=4$.}
\label{fig7}
\end{figure}

In Fig.~\ref{fig7} we compare the temperature dependence of the compressibility of a system in the presence of spin-independent and spin-dependent disorder, respectively. In both cases disorder reduces the compressibility and, as in the case of the DOS, the effect of spin-independent disorder is stronger. We also observe an enhancement of the statistical fluctuations within the Monte Carlo method at stronger interactions.

Finally we note that at any given temperature $T$ the reduction of the compressibility $\chi_c$ by disorder is smaller in the strongly correlated case than in the weakly correlated case. This holds true for both types of disorder, and is due to the fact that a strong repulsive interaction leads to a rigidity of the system which makes it less sensitive to the influence of disorder.

The non-interacting model discussed in Section \ref{DMFT} where particles with different spin have different bandwidths \cite{Skolimowski} yields qualitatively similar results for the ferromagnetic susceptibility and the compressibility.
Although such a toy model cannot explain the details of the magnetization it appears to be quite useful for explaining qualitative features of systems with spin-dependent disorder.

\section{Spin-imbalanced fermions}

\begin{figure}[!]
\centerline{\includegraphics[width=15cm]{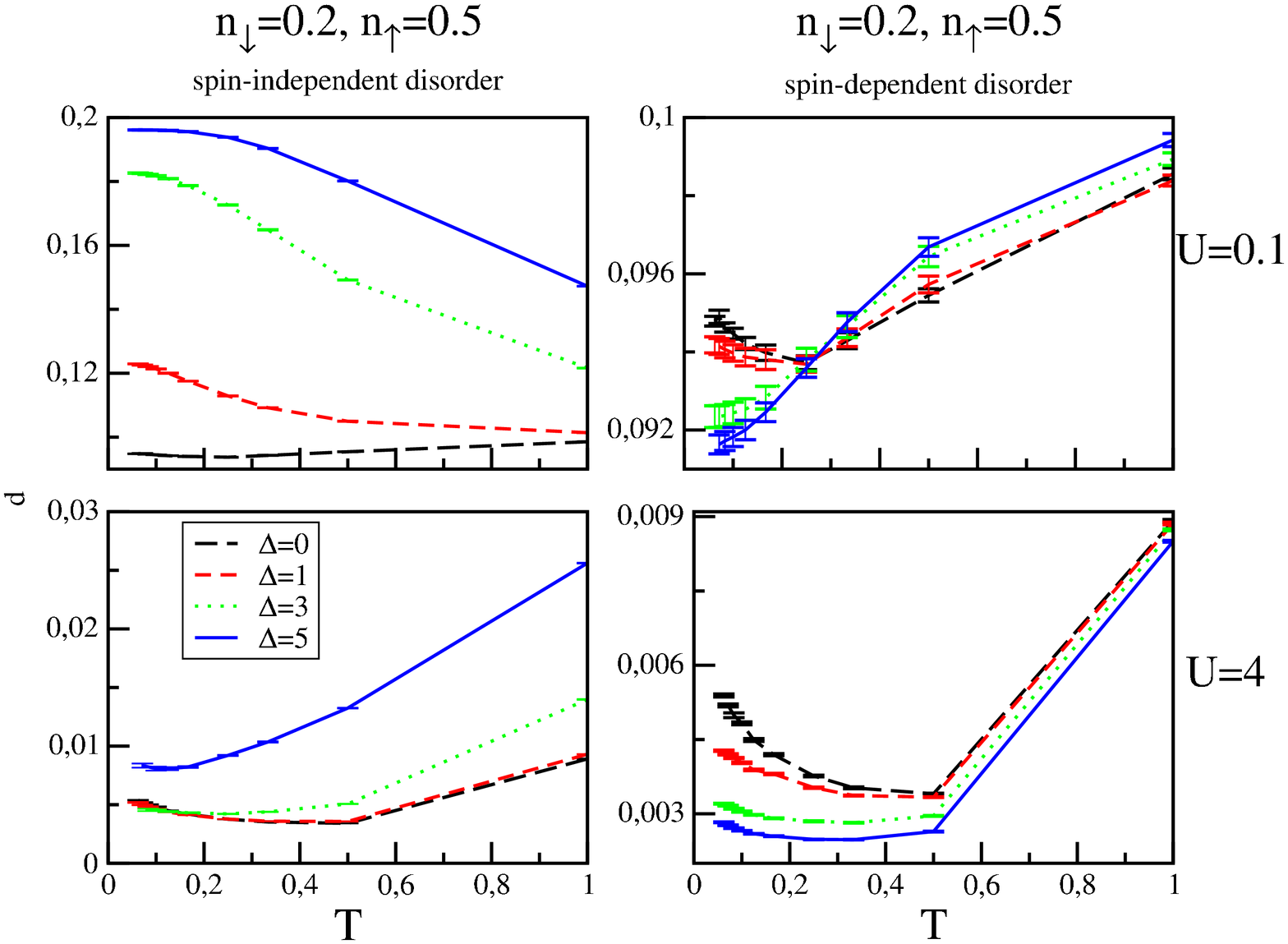}}
\caption{Spin-imbalanced fermions: Double occupation $d$ as a function of the temperature $T$ for different values of $\Delta=0,1,3,5$ for spin-independent  disorder (left column) and spin-dependent disorder (right columns); upper panels: $U=0.1$,  lower panels: $U=4$. The density of particles  is $n_{\downarrow}=0.2$ and $n_{\uparrow}=0.5$. }
\label{fig8}
\end{figure}

In the case of spin-imbalanced fermions the number of fermions with spin $\sigma$ is conserved individually. The magnetization  $m = n_{\uparrow} - n_{\downarrow}$ is then constant, i.e., does not depend on the thermodynamic variables and parameters of the model. In Fig.~\ref{fig8} we compare the temperature dependence of the average double occupation $d$ of the spin-imbalanced fermions for spin-independent disorder (left column) and spin-dependent disorder (right columns), respectively. For spin-independent disorder in the presence of a weak interaction ($U=0.1$) an increase of the disorder strength $\Delta$  leads to a corresponding increase of $d$, i.e., on average more energy levels below the Fermi energy are occupied by two fermions with opposite spins.  For increasing $U$ a minimum appears in $d(T)$ which corresponds to a maximum in the local moment. This is seen clearly for $U=4$ in the left column of Fig.~\ref{fig8}. Therefore a maximum of the of local moments is found in both models considered here. In the case of spin-dependent disorder (right columns of Fig.~\ref{fig8}) the double occupation is, in general, suppressed  by the disorder $\Delta$.  This is an effect of the finite magnetization of the system, which is here imposed by fixing the individual spin densities. As a consequence the local moment increases. We also see that the local minimum of $d(T)$ is shifted to lower temperatures when $\Delta$ is increased. Whether it disappears at $T=0$ cannot be decided here since, as mentioned earlier,  we are not able to run QMC simulations at  lower temperatures. It is interesting to note that the behavior of the double occupation is qualitatively similar in both models.

\begin{figure}[!]
\centerline{\includegraphics[width=15cm]{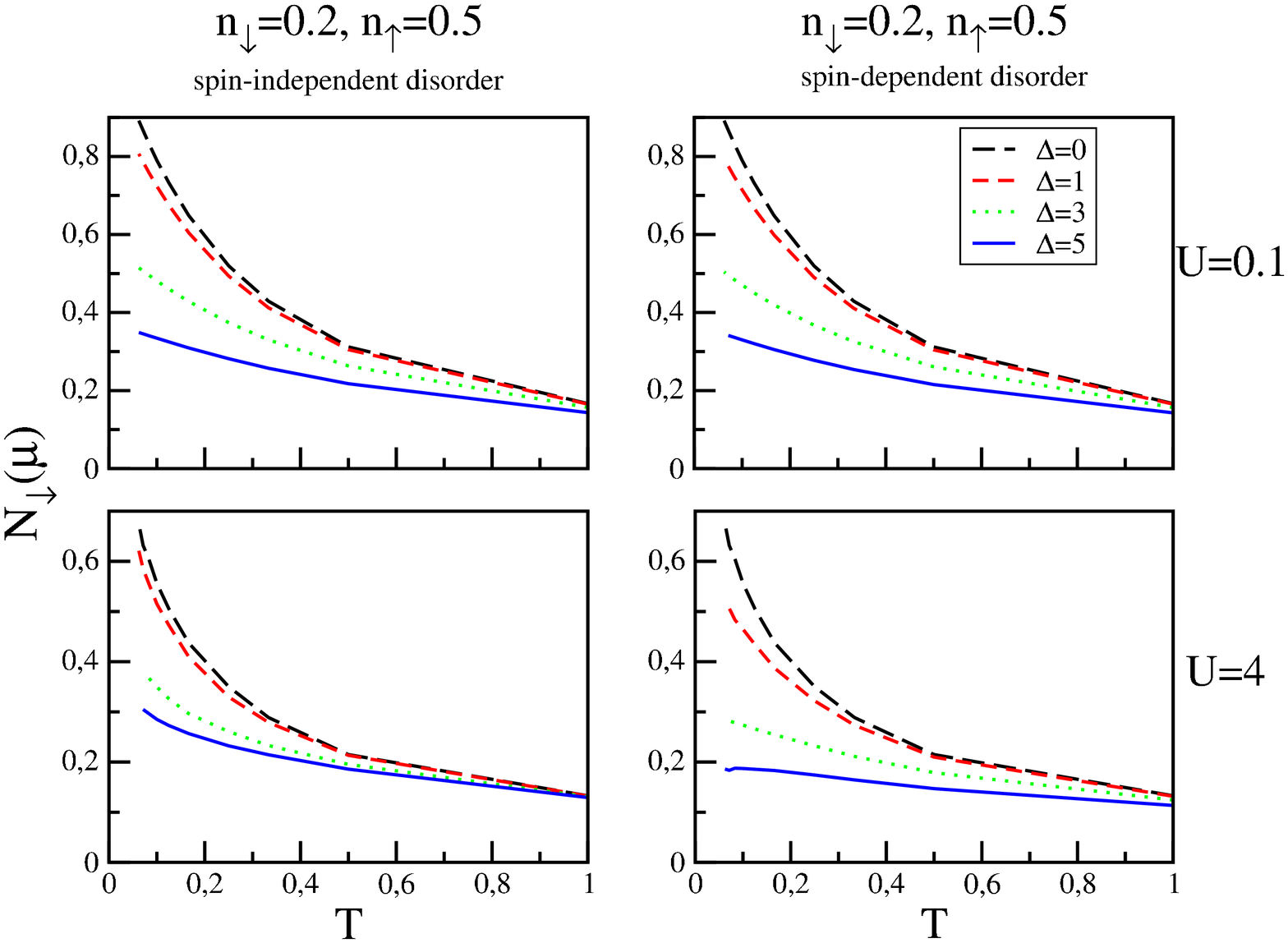}}
\centerline{\includegraphics[width=15cm]{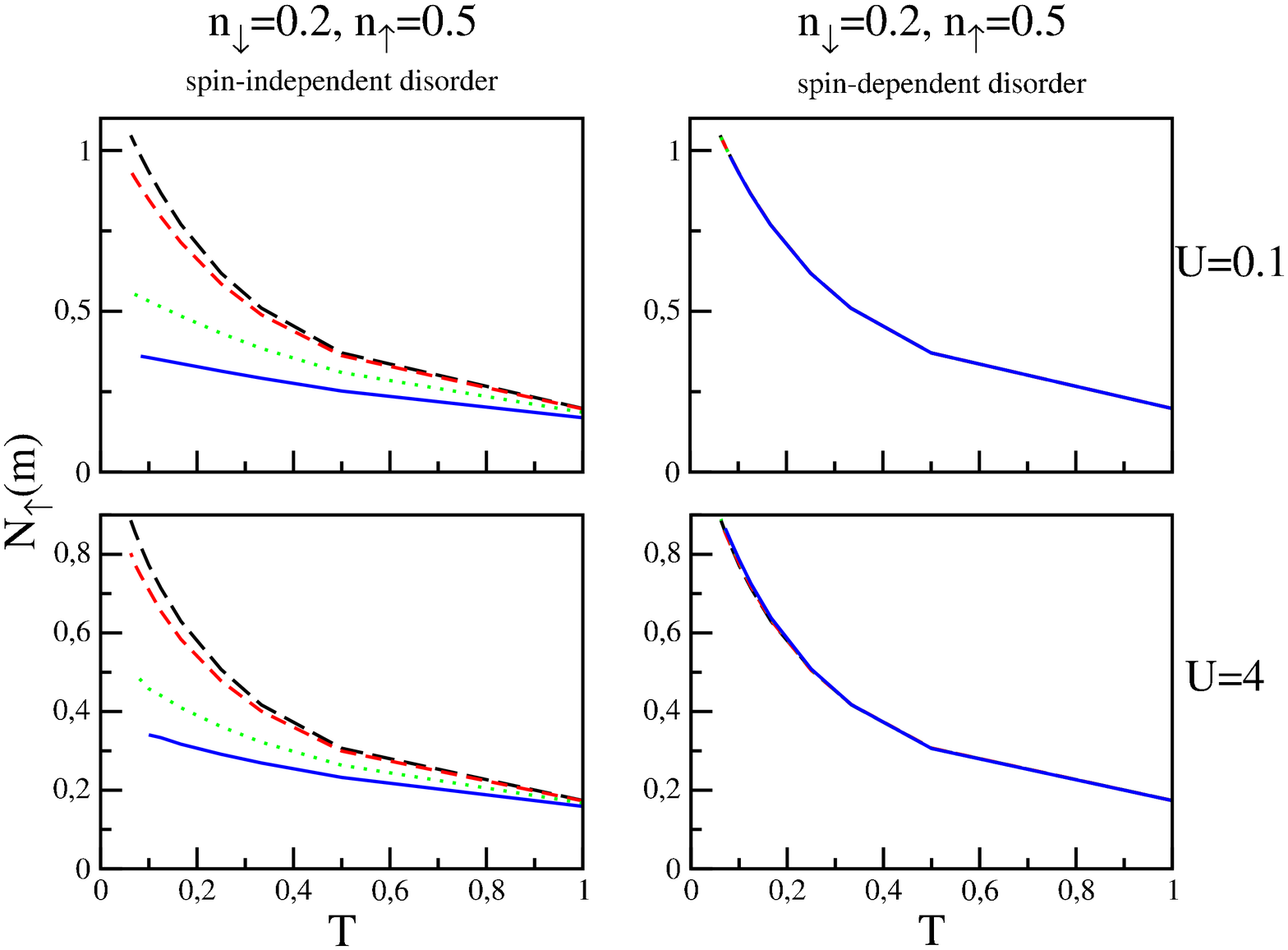}}
\caption{Spin-imbalanced fermions: Spin resolved density of states for spin down (upper figure) and spin up (lower figure)  at the chemical potentials as a function of the temperature $T$ for different values ($\Delta=0,1,3,5$) of the spin-independent  disorder disorder (left) and the spin-dependent disorder (right); upper panels: $U=0.1$, lower panels: $U=4$.
The density of particles  is $n_{\downarrow}=0.2$ and $n_{\uparrow}=0.5$.}
\label{fig9}
\end{figure}

In Fig.~\ref{fig9} we compare the values of the spin resolved DOS at the chemical potential for different types of disorder and different disorder strengths. Spin-independent disorder always reduces the DOS, while spin-dependent disorder only reduces the DOS of the corresponding spin subsystem. For weak interactions the DOS of the opposite spin subsystem is almost unchanged. Only at stronger interactions ($U=4$) does the disorder in one spin subsystem also influence the DOS of the opposite spin subsystem (but only weakly). This correlation effect is expected, since the two spin subsystems are coupled by the on-site interaction $U$.

In the case of spin-imbalanced fermions it is useful to discuss a compressibility \emph{matrix} \cite{Seo11}
\begin{equation}
\chi_{c\;\sigma \sigma'}  = \left(\frac{\partial n_{\sigma}(\mu_{\uparrow},\mu_{\downarrow},T)}{\partial \mu_{\sigma '}}\right)_T = \beta \langle \langle \hat{n}_{\sigma} \hat{n}_{\sigma '}\rangle - \langle \hat{n}_{\sigma}\rangle \langle  \hat{n}_{\sigma '}\rangle \rangle _{\rm dis}/N_L,
\end{equation}
where $\hat{n}_{\sigma} = \sum_i \hat{n}_{i\sigma}$.
Its off-diagonal elements provide a measure of density-density correlations between different spin subsystems. We note that the matrix is symmetric $ \chi_{c\; \uparrow \downarrow }  = \chi_{c \;\downarrow \uparrow }  $.
At $U=0$ one has $\chi_{c \;\uparrow \downarrow }  =0$.

\begin{figure}[!]
\centerline{\includegraphics[width=15cm]{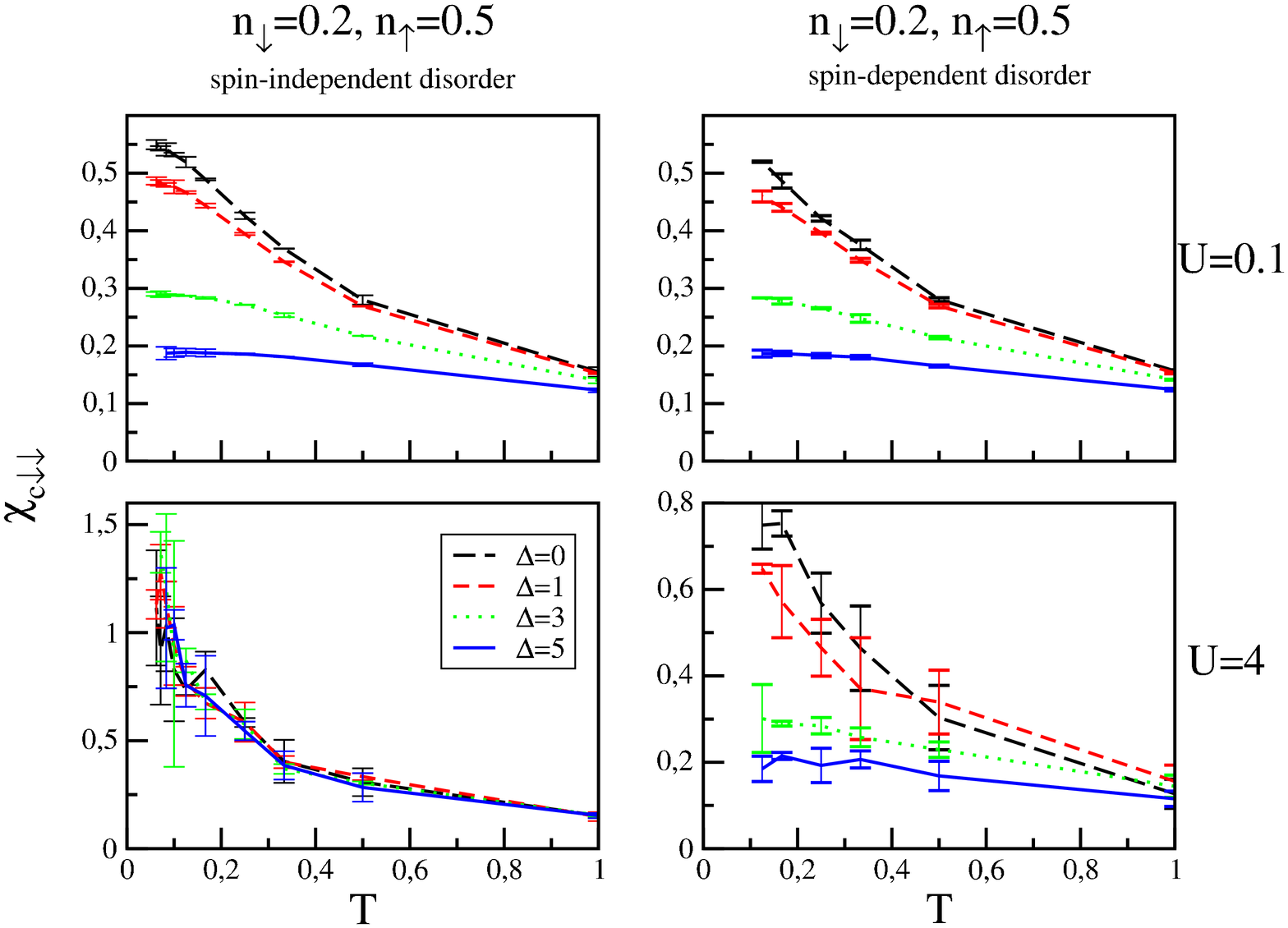}}
\centerline{\includegraphics[width=15cm]{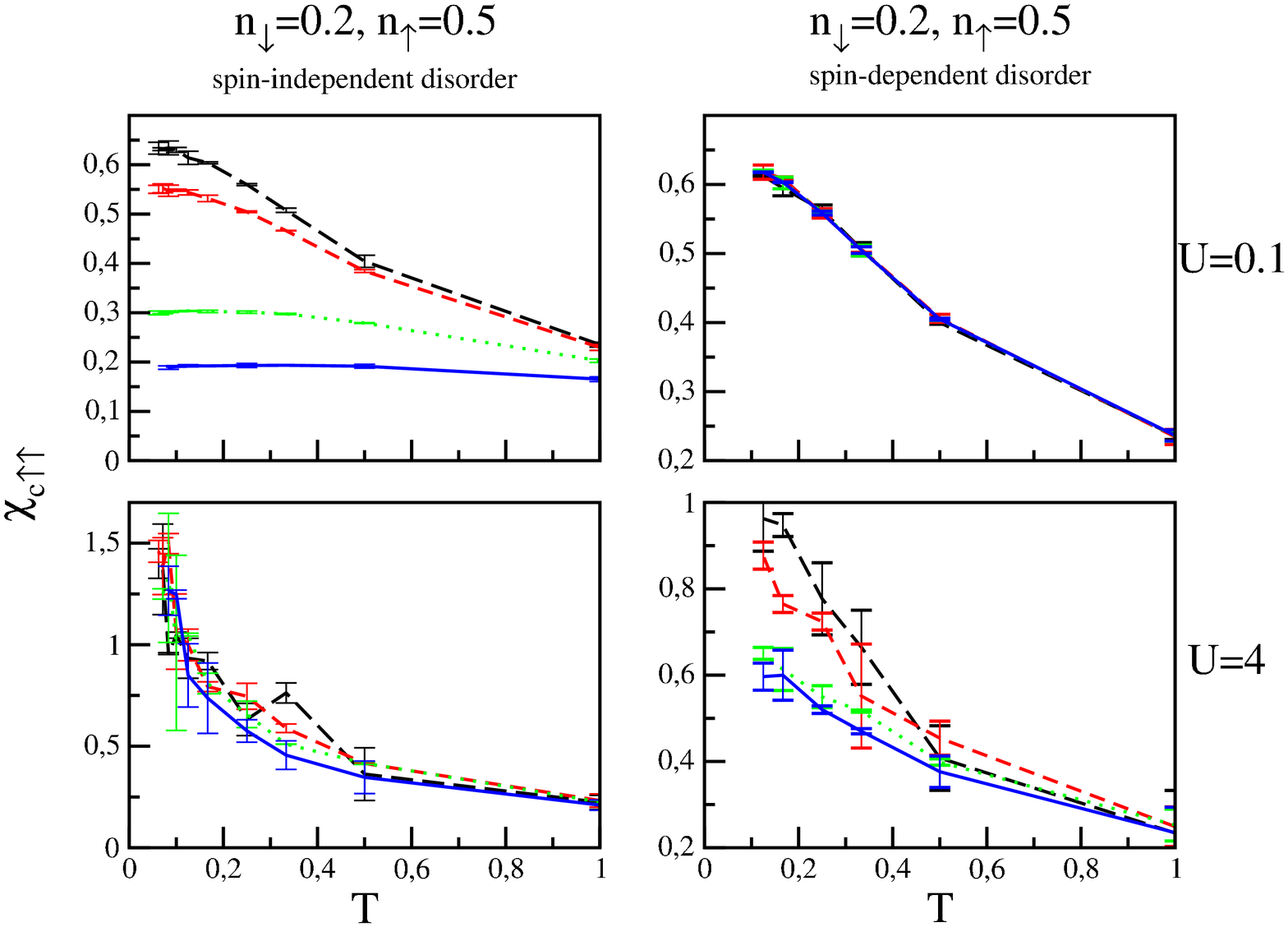}}
\caption{Spin-imbalanced fermions: Diagonal compressibilities $ \chi_{c \;\downarrow \downarrow }$ (upper figure) and $\chi_{c \;\uparrow \uparrow }$ (lower figure) as a function of the temperature $T$ for different values ($\Delta=0,1,3,5$) of spin-independent disorder (left columns) and spin-dependent disorder (right columns); upper panels: $U=0.1$,  lower panels: $U=4$. The density of particles  is $n_{\downarrow}=0.2$ and $n_{\uparrow}=0.5$.}
\label{fig10}
\end{figure}

In Fig.~\ref{fig10} the diagonal compressibilities in different spin channels are shown. In the case of spin-independent  disorder (left columns in Fig.~\ref{fig10}) the diagonal compressibilities are found to be reduced in both spin subbands for increasing disorder strength $\Delta$. Due to the imbalance of the spin population one has $\chi_{c \uparrow \uparrow } \neq \chi_{c \downarrow \downarrow } $.

In the case of spin-dependent disorder the spin-down compressibility $\chi_{c \;\downarrow \downarrow } $ is reduced when the disorder $\Delta$  is increased; this holds for any interaction strength $U$, cf. upper figure, right columns in Fig.~\ref{fig10}. By contrast, the spin-up compressibility $\chi_{c \;\uparrow \uparrow }$  remains almost unchanged at weak interactions. But for larger $U$ it decreases for increasing disorder $\Delta$, cf. lower figure, right columns  in Fig.~\ref{fig10}.
The disorder-induced changes in $ \chi_{c \;\uparrow \uparrow }$ are an interaction effect, since the disorder $\Delta$ acts only on the spin-down subsystem.

For spin-imbalanced fermions  we observe a  distinctive qualitative difference between spin-independent and spin-dependent disorder. Namely, as for the model where only the total number of fermions is kept constant,  we find that  at any given temperature $T$ the reduction of the compressibility $\chi_c$ by spin-independent disorder is less in the strongly correlated case than in the weakly correlated case (quantitatively this effect is  even much stronger here, cf. Fig.~\ref{fig7}). But the opposite happens in the case of spin-dependent disorder: then the reduction of the compressibility by disorder increases with increasing interaction strength $U$.

\begin{figure}[!]
\centerline{\includegraphics[width=15cm]{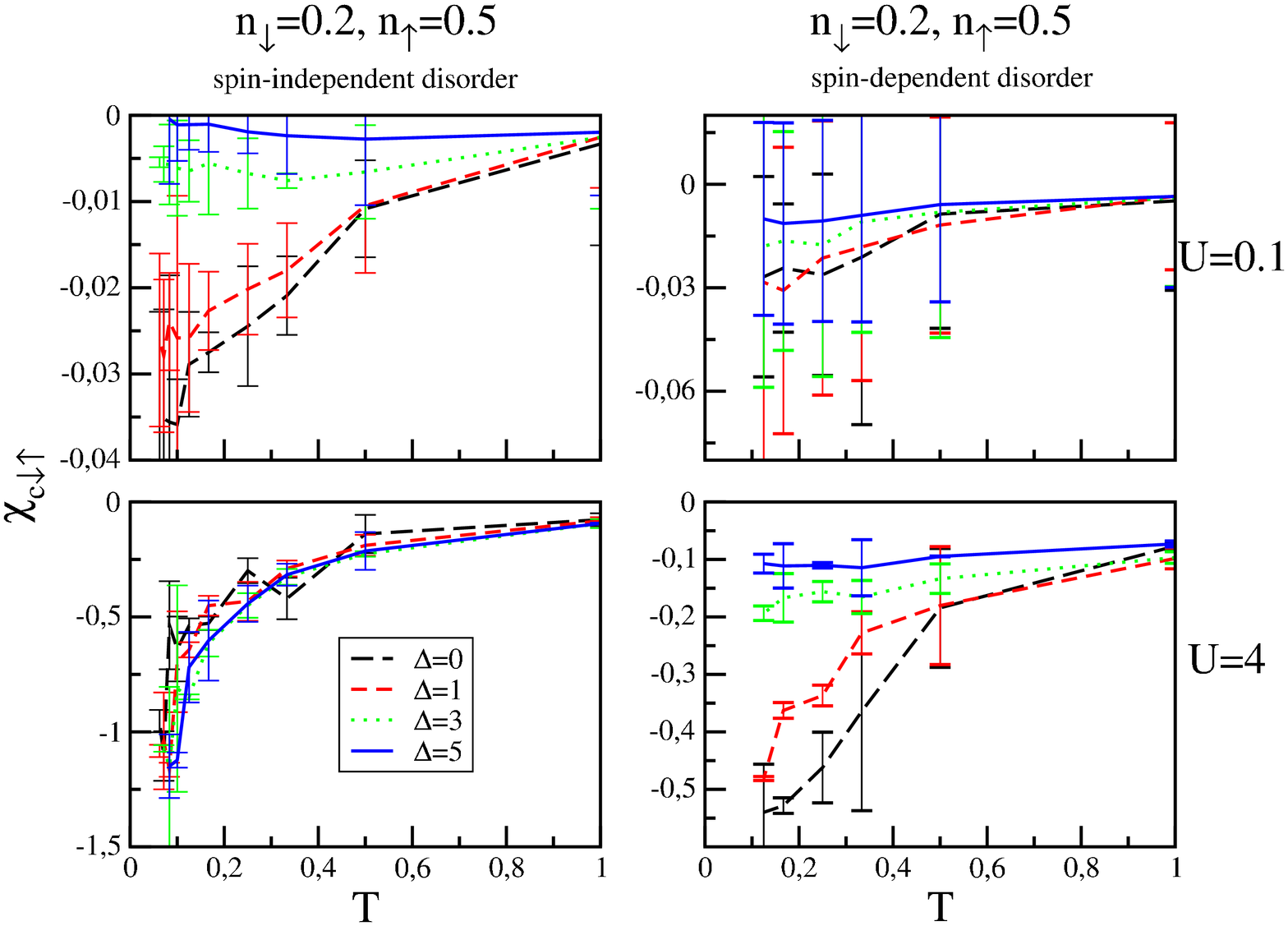}}
\caption{Spin-imbalanced fermions: Off-diagonal compressibilities $ \chi_{c \;\downarrow \uparrow }$ as a function of the temperature $T$ for different values ($\Delta=0,1,3,5$) of spin-independent disorder (left columns) and spin-dependent disorder (right columns); upper panels: $U=0.1$,  lower panels: $U=4$. The density of particles  is $n_{\downarrow}=0.2$ and $n_{\uparrow}=0.5$. Note that the relative error bars are of the same order in all cases but the off-diagonal susceptibility is very small at weak interactions, cf. compare scales on vertical axis.}
\label{fig11}
\end{figure}

In Fig.~\ref{fig11} we show the off-diagonal elements of the compressibility. As expected their absolute values increase with increasing interaction $U$. We also observe that for both types of disorder an increase of the disorder strength leads to a reduction of the inter-spin correlations. Although the off-diagonal elements of the compressibility are negative the determinant of the compressibility matrix is positive. Therefore, the system is stable against phase separation. The effect of disorder on the off-diagonal compressibilities at a given temperature $T$ is similar to that of the diagonal compressibilities.

\section{Conclusions and Outlook}

In summary, we explored the thermodynamic properties of correlated fermions in the presence of spin-independent and spin-dependent disorder within dynamical mean-field theory. We discussed two models where either the total number of fermions, or the number of each spin component, is conserved. These two cases can be realized in experiments on cold atoms in optical lattices.

In the first model we found that, in contrast to spin-independent disorder, spin-\emph{dependent} disorder induces a ferromagnetic polarization. However, instead of a Zeeman-Stoner-type shift of the sub-bands, which would be observed in the presence of an external magnetic field, a spin-dependent broadening takes place.

In the second model we showed that disorder which acts only on fermions with one particular spin direction nevertheless  also affects the properties of particles with the opposite spin direction; this effect is more easily observed at strong interactions. Therefore, the investigation of the  spin-resolved
densities of states and spin-resolved compressibilities will provide useful quantitative information about
correlations between the two spin subsystems.

\bigskip

We thank R. Scalettar for drawing our attention to this problem and for helpful discussions. We also thank M. Jiang for sharing with us his notes on related investigations. Discussions with L. Chioncel and J. Kune\v{s} are gratefully acknowledged. K.M, J.S, and K.B. acknowledge support by the Foundation for Polish Science (FNP) through the TEAM/2010-6/2 project, co-financed by the EU European Regional Development Fund. This research was also supported in part by the Deutsche Forschungsgemeinschaft through TRR 80 (PBC, KB, DV).


\section*{References}
\begin{thebibliography}{100}


\bibitem{Mott90} N. F. Mott, \emph{Metal-Insulator Transitions} (Taylor and Francis, London, 1990).

\bibitem{Lee85} P. A. Lee and T. V. Ramakrishnan, Rev. Mod. Phys. {\bf 57}, 287 (1985).

\bibitem{Altshuler85}  B. L. Altshuler and A. G. Aronov, in \emph{Electron-Electron Interactions in Disordered Systems}, ed. M. Pollak and A. L. Efros (North-Holland, Amsterdam, 1985), p. 1.

\bibitem{Belitz94} D. Belitz and T. R. Kirkpatrick, Rev. Mod. Phys. {\bf 66}, 261 (1994).

\bibitem{Abrahams01}    E. Abrahams, S. V. Kravchenko, and M. P. Sarachik, Rev. Mod. Phys. {\bf 73}, 251 (2001).

\bibitem{Abrahams10} \emph{50 Years of Anderson Localization}, ed. E. Abrahams (World Scientific, Singapore 2010).


\bibitem{Lewenstein07}    M. Lewenstein, A. Sanpera, V. Ahufinger, B. Damski, A. Sen, and U. Sen, Adv. Phys. {\bf 56}, 243 (2007).

\bibitem{Fallani07}    L. Fallani, J. Lye, V. Guarrera, C. Fort, and M. Inguscio, Phys. Rev. Lett. {\bf 98}, 130404 (2007).


\bibitem{Billy08}  J. Billy, V. Josse, Z. Zuo, A. Bernard, B. Hambrecht, P. Lugan, D. Clement, L. Sanchez-Palencia, P. Bouyer, and A. Aspect, Nature (London) {\bf 453}, 891 (2008).


\bibitem{Roati08}  G. Roati, C. D Errico, L. Fallani, M. Fattori, C. Fort, M. Zaccanti, G. Modugno, M. Modugno, and   M. Inguscio, Nature (London) {\bf 453}, 895 (2008).


\bibitem{White09}  M. White, M. Pasienski, D. McKay, S. Q. Zhou, D. Ceperley, and B. DeMarco, Phys. Rev. Lett. {\bf 102}, 055301 (2009).


\bibitem{Lewenstein2010} L. S.-Palencia and M. Lewenstein, Nat. Phys. {\bf 6}, 87 (2010).

\bibitem{Kondov2011} S. S. Kondov, W. R. McGehee, J. J. Zirbel, and B. De Marco, Science 334, 66 (2011).

\bibitem{Jendrzejewski12} F. Jendrzejewski, A. Bernard, K. Mueller, P. Cheinet, V. Josse, M. Piraud, L. Pezz\'{e}, L. Sanchez-Palencia, A. Aspect, and P. Bouyer, Nat. Phys. {\bf 8}, 398 (2012).


\bibitem{Mandel03} O. Mandel, M. Greiner, A. Widera, T. Rom, Th. W. H\"{a}nsch, and I. Bloch, Phys. Rev. Lett. {\bf 91}, 010407 (2003).

\bibitem{Mckay10} D. McKay and B. DeMarco, New J. Phys. {\bf 12},  055013 (2010).

\bibitem{Soltan11}  P. Soltan-Panahi,
 J. Struck, P. Hauke,  A. Bick,  W. Plenkers,  G. Meineke,  C. Becker,  P. Windpassinger,  M. Lewenstein,  and K. Sengstock, Nat. Phys. {\bf 7}, 434 (2011).

\bibitem{Liu04} W. V. Liu, F. Wilczek, and P. Zoller, Phys. Rev. A {\bf 70}, 033603 (2004).

\bibitem{Cazalilla05} M. A. Cazalilla, A. F. Ho, and T. Giamarchi, Phys. Rev. Lett. 95, 226402 (2005).

\bibitem{Feiguin09} A. E. Feiguin and M. P. A. Fisher, Phys. Rev. Lett. 103, 025303 (2009).

\bibitem{FK-model} In principle, the Falicov-Kimball model \cite{FK,FK-RMP} also belongs into this class of models since it may be viewed as a fermionic lattice model with spin-dependent hopping amplitudes. Namely, one spin species can hop, while the other one is fixed to the lattice.

\bibitem{FK} L. M. Falicov and J. C. Kimball, Phys. Rev. Lett. \textbf{22}, 997 (1969).

\bibitem{FK-RMP} J. K. Freericks and V. Zlati\'{c}, Rev. Mod. Phys. \textbf{75}, 1333 (2003).

\bibitem{Scalettar12} R. Nanguneri, M. Jiang, T. Cary, G. G. Batrouni, and R. T. Scalettar, Phys. Rev. B {\bf 85}, 134506 (2012).

\bibitem{Pasienski2010} M. Pasienski, D. McKay, M. White, and B. DeMarco, Nat. Phys. 6, 677 (2010).

\bibitem{mott68}
N.~F. Mott, Rev.\ Mod.\ Phys. \textbf{40}, 677 (1968).


\bibitem{Gebhard}
F.~Gebhard, \emph{The Mott Metal-Insulator Transition}, Springer, Berlin, 1997.

\bibitem{Georges}
A.~Georges, G.~Kotliar, W.~Krauth, and M.~J. Rozenberg, Rev.\ Mod.\
  Phys. \textbf{68}, 13 (1996).

\bibitem{tokura}
M.~Imada, A.~Fujimori, and Y.~Tokura, Rev.\ Mod.\ Phys. \textbf{70},
  1039 (1998).


\bibitem{dmft_phys_today}
G.~Kotliar and D.~Vollhardt, Physics Today \textbf{3}, 53 (2004).

\bibitem{MIT-off-half-filling} The statement that a Mott-Hubbard metal-insulator transition can take place only at half filling is correct only in the case of continuous probability distributions of the disorder as employed in our paper. Indeed, disorder in a binary alloy
    $A_{x}B_{1-x}$ with  a bimodal probability distribution leads to a band splitting at sufficiently strong disorder, giving rise to alloy subbands. For filling factors $\nu =x$ or $1+x$ the lower or upper alloy subband is then half filled
and the system becomes a Mott insulator at strong interactions, with a
correlation gap at the Fermi level. In this case a  Mott-Hubbard transition takes place \emph{off half-filling} \cite{MIT-off-half-filling_ref2}.

%\bibitem{MIT-off-half-filling_ref1} K. Byczuk, M. Ulmke, and D. Vollhardt, Phys. Rev. Lett. {\bf 90}, 196403 (2003).

\bibitem{MIT-off-half-filling_ref2} K. Byczuk, W. Hofstetter, and D. Vollhardt, Phys. Rev. B {\bf 69}, 045112 (2004).

\bibitem{Gonis} A. Gonis, {\it Green functions for ordered and disordered systems}, (North-Holland, 1992).

\bibitem{Byczuk05} K. Byczuk, W. Hofstetter, and D. Vollhardt, Phys. Rev. Lett. {\bf 94}, 056404 (2005).

\bibitem{Byczuk09} K. Byczuk, W. Hofstetter, and D. Vollhardt, Phys. Rev. Lett. {\bf 102}, 146403 (2009).

%\bibitem{Byczuk2009correlated} K. Byczuk, W. Hofstetter, U. Yu, and D. Vollhardt, Eur. Phys. J. Special Topics {\bf 180}, 135 (2010).

\bibitem{Chakraborty11} P. B. Chakraborty, K Byczuk, and D. Vollhardt, Phys. Rev. B {\bf 84}, 035121 (2011).


\bibitem{Dobrosavljevic12} V. Dobrosavljevic, {\it Introduction to Metal--Insulator Transitions},
in "Conductor--Insulator Quantum Phase Transitions", edited by V. Dobrosavljevic, N. Trivedi, and J.M. Valles Jr (Oxford University Press, 2012) pp 3--58.

\bibitem{Miranda12} E. Miranda and V. Dobrosavljevic, {\it Dynamical mean--field theories of correlation and disorder},
in "Conductor--Insulator Quantum Phase Transitions", edited by V. Dobrosavljevic, N. Trivedi, and J.M. Valles Jr (Oxford University Press, 2012) pp 161--236.

\bibitem{Fattori06} M. Fattori, T. Koch, S. Goetz, A. Griesmaier, S. Hensler, J. Stuhler and T. Pfau, Nat. Phys. {\bf 2}, 765 (2006).
p. 339;
\bibitem{Ketterle13} Lee Y-R, Wang T T, Rvachov T M, Choi J-H, Ketterle W and Heo M-S, arXiv:1301.1139.

%

\bibitem{Zwerlein06} M. W. Zwierlein, A. Schirotzek, Ch. H. Schunck, W. Ketterle, Science {\bf 311}, 492 (2006).

\bibitem{Shin06} Y. Shin, M. W. Zwierlein, C. H. Schunck, A. Schirotzek, W. Ketterle, Phys. Rev. Lett. {\bf 97}, 030401 (2006).

\bibitem{Navon10} N. Navon, S. Nascimbene, F. Chevy, Ch. Salomon, Science {\bf 328}, 729 (2010).

\bibitem{Parish07} M. M. Parish, F. M. Marchetti, A. Lamacraft, B. D. Simons, Nat. Phys. {\bf 3}, 124 (2007).

\bibitem{Wunsch10} B. Wunsch, L. Fritz, N.T. Zinner, E. Manousakis, E. Demler, Phys. Rev. A {\bf 81}, 013616 (2010).

\bibitem{Snoek11} M. Snoek, I. Titvinidze, W. Hofstetter, Phys. Rev. B. {\bf 83}, 054419 (2011).

\bibitem{Gubbels12} K. B. Gubbels, H. T. C. Stoof, arXiv:1205.0568.

\bibitem{Wolak12} M. J. Wolak, B. Gr{\'e}maud, R. T. Scalettar, G. G. Batrouni, arXiv:1206.5050.

\bibitem{Lloyd} P. Lloyd, J. Phys. C {\bf 2}, 1717 (1969).


\bibitem{Thouless} D. Thouless, Phys. Rep. {\bf 13}, 93 (1974).

\bibitem{Wegner} F. Wegner, Z. Phys. B {\bf 44}, 9 (1981).


\bibitem{Scalettar_private} We thank Richard Scalettar for discussions concerning this interpretation.

\bibitem{Byczuk-review} K. Byczuk, W. Hofstetter und D. Vollhardt in \emph{50 Years of Anderson Localization}, ed. E. Abrahams, p. 473 (World Scientific, Singapore, 2010); reprinted in Int. J. Mod. Phys. B 24, 1727 (2010).

\bibitem{Metzner} W. Metzner and D. Vollhardt, Phys. Rev. Lett. {\bf 62}, 324 (1989).

\bibitem{Ulmke} M. Ulmke, V. Jani\v{s}, and D. Vollhardt, Phys. Rev. B {\bf 51}, 10411 (1995).

\bibitem{Janis91} V. Jani\v{s}, Z. Phys. B \textbf{83}, 227 (1991).

\bibitem{Vlaming} R. Vlaming and D. Vollhardt, Phys. Rev. B {\bf 45}, 4637 (1992).

\bibitem{Vollhardt04} D.~Vollhardt, in \emph{Lectures on the Physics of Strongly Correlated Systems {XIV}}, \emph{AIP Conference Proceedings}, vol. 1297, ed. by A.~Avella,  F.~Mancini (American Institute of Physics, Melville, 2010), p. 339; arXiv:1004.5069v3.

\bibitem{ulmke98} M.\ Ulmke, Eur.\ Phys.\ J.\ {\bf B 1}, 301 (1998).

\bibitem{wahle98} J.\ Wahle, N. Bl\"{u}mer, J. Schlipf, K. Held, and D. Vollhardt,
Phys.\ Rev.\ B {\bf 58}, 12749 (1998).

\bibitem{hirsh86} J.\ E.\ Hirsch and  R.\ M.\ Fye, Phys.\ Rev.\ Lett.\ {\bf 56}, 2521 (1986).

\bibitem{error-bars} The error bars shown in Figs. \ref{fig2}, \ref{fig1}, \ref{fig3}, \ref{fig4}, \ref{fig6}, \ref{fig7}, \ref{fig8} and \ref{fig9} indicate the uncertainties due to Monte Carlo sampling and are determined by the standard mean deviations. In Figs. \ref{fig1},  \ref{fig4} and \ref{fig9} the error bars are smaller than the line width and are therefore not visible. In Figs. \ref{fig5}, \ref{fig7}, \ref{fig10} and \ref{fig11} the error bars are obtained by the linear regression method since the derivatives were obtained by fitting straight lines.

\bibitem{Skolimowski} J. Skolimowski, {\it Termodynamika fermion\'ow w obecno\'sci nieporzadku zale\.znego od spinu}, Diploma Thesis, University of Warsaw (2012), (unpublished) (in Polish)

\bibitem{Hirsch} J. E. Hirsch, Phys. Rev. B {\bf 59}, 6256 (1999).

\bibitem{Kim} K. Kim, U. Yu, B. H. Kim, and B. I. Min, J. Phys.: Cond. Mat. {\bf 18}, 7227 (2006).

\bibitem{Chakraborthy11_bis} P. B. Chakraborty, K Byczuk, and D. Vollhardt,
Phys. Rev. B {\bf 84}, 155123 (2011).

\bibitem{Seo11} Kangjun Seo, C. A. R. S{\'a}de Melo, arXiv:1101.361; Kangjun Seo, C. A. R. S{\'a}de Melo, arXiv:1105.4365.

\end{thebibliography}
\end{document}